\documentclass[prd,nofootinbib,preprintnumbers,preprint]{revtex4}
\usepackage{amsmath}
\usepackage{amsfonts}
\usepackage{amssymb}
\usepackage{dsfont}
\usepackage[hyperfootnotes=false]{hyperref}
\usepackage[dvipsnames]{xcolor}

\newcommand{\mX}{\mathcal X}
\newcommand{\mY}{\mathcal Y}
\newcommand{\mZ}{\mathcal Z}
\newcommand{\mK}{\mathcal K}

\newcommand{\mF}{\mathcal F}
\newcommand{\mG}{\mathcal G}
\def\e{{\rm e}}

\newcommand{\RP}{\right\rangle}
\newcommand{\LP}{\left\langle}
\newcommand{\beq}{\begin{equation}}
\newcommand{\eeq}{\end{equation}}
\newcommand{\bAL}{\begin{align}}
\newcommand{\eAL}{\end{align}}

\newcommand{\LCM}{\nabla}

\newcommand{\n}{\nonumber}

\newcounter{example}[section]
\newcounter{remark}[section]
\newcounter{theorem}[section]
\newcounter{proposition}[section]
\newcounter{lemma}[section]
\newcounter{corollary}[section]
\newcounter{definition}[section]

\def\theremark{\arabic{section}.\arabic{remark}}
\def\thetheorem{\arabic{section}.\arabic{theorem}}

\def\thedefinition{\arabic{section}.\arabic{definition}}

\renewcommand*{\email}[1]{\footnote{Electronic address: \href{mailto:#1}{\nolinkurl{#1}} }}

\newenvironment{theorem}{\refstepcounter{theorem}
\medskip\noindent{\bf Theorem \thetheorem}:}{$\Box$\medskip}
\newenvironment{proposition}{\refstepcounter{theorem}\medskip\noindent{\bf
Proposition \thetheorem}:}{$\Box$\medskip}

\newenvironment{definition}{\refstepcounter{definition}\medskip\noindent{\bf
Definition \thedefinition}:}{$\Box$\medskip}

\begin{document}

\title{Killing tensors in foliated spacetimes and photon surfaces}

\author{Kirill Kobialko\email{kobyalkokv@yandex.ru}}
\author{Igor Bogush\email{igbogush@gmail.com}}
\author{Dmitri Gal'tsov\email{galtsov@phys.msu.ru}}
\affiliation{Faculty of Physics, Moscow State University, 119899, Moscow, Russia}

\begin{abstract}
We discuss a recently proposed geometric method \cite{Kobialko:2021aqg} for constructing a nontrivial Killing tensor of rank two in a foliated spacetime of codimension one that lifts trivial Killing tensors from slices to the entire manifold. The existence of nontrivial Killing tensor is closely related to generalized photon  surfaces. The method is illustrated on  some known cases and  used to construct the hitherto unknown Killing tensor for the Nutty dyon in dilaton-axion gravity.
\end{abstract}

\maketitle

\setcounter{page}{2}

\setcounter{equation}{0}
\setcounter{subsection}{0}
\setcounter{section}{0}

\section{Introduction}\label{intro}
Killing tensors of the second rank express hidden symmetries of spacetime \cite{Carter:1977pq,Frolov:2017kze}  providing  integrals of motion for geodesics and wave operators in field theories. Some constructive procedures to find them were suggested for spaces with a warped/twisted product structure \cite{Krtous:2015ona}, spaces admitting a hypersurface orthogonal Killing vector field \cite{Garfinkle:2010er,Garfinkle:2013cha}, or special conformal Killing vector fields \cite{Rani:2003br}. Recently, some deformed Kerr metrics attracted attention when trying to find new physics in ultracompact astrophysical objects.
Classes of such metrics were listed that admit Killing tensors
\cite{Carson:2020dez,Papadopoulos:2020kxu}, but for others the separation of variables in geodesic equations is not guaranteed.

The new procedure \cite{Kobialko:2021aqg} suggested for foliated spacetime of codimension one is based on lifting the trivial Killing tensors in slices with an arbitrary second fundamental form \cite{Kobialko:2021aqg}. This remove some particular assumptions about slices, so, hopefully, we can move further in the study of separability.
The method is closely related to formalism of
fundamental photon surfaces introduced in \cite{Kobialko:2020vqf}. This  generalizes one previous observation
\cite{Pappas:2018opz} on the relationship between spacetime separability and spherical photon orbits. It is also worth noting that our method  does not require sovling diffenerial equations at all.

As illustrations, we apply this new technique to some conventional metrics of Petrov type D demonstrating that this technique allows to obtain known Killing tensors \cite{Kubiznak:2007kh,Vasudevan:2005bz} purely algebraically, without solving any differential equations. Then we successfully apply new technique to find Killing tensor for Gal'tsov-Kechkin solution \cite{Galtsov:1994pd,Garcia:1995qz} in dilaton-axion gravity which belongs to the general Petrov type I. The new method reveals the nature of Killing tensors as arising from isometries of low-dimensional slices of a smooth foliation.

In Sec. \ref{sec:killing} we briefly describe the equations for the Killing vectors and Killing tensors of rank two. In Sec. \ref{sec:generation} we consider spacetimes with foliation of codimension one and present the equations governing the interplay between symmetries in the bulk and in the slices. Eventually we describe the Killing generating technique. In Sec. \ref{sec:photon} we reveal the connection between the Killing tensors and the fundamental photon surfaces. Sec. \ref{sec:axially} provides examples with axial symmetry in different models. 

\section{Conventions}
\label{sec:killing}

Let $M$ be a Lorentzian manifold of dimension $m$ with scalar product $\LP \;\cdot\; ,\;\cdot\;\RP$ and Levi-Civita connection $\LCM$\footnote{Here we also use the notation $\underset{\mX \leftrightarrow \mY}{{\rm Sym}}\left\{ B(\mX, \mY) \right\}\equiv B(\mX, \mY)+ B(\mY, \mX)$.}.

\begin{definition} 
A vector field $\mK:M\rightarrow TM$ is called a Killing vector field if \cite{Chen} 
\begin{align}\label{eq:killing_equation}
\underset{\mX \leftrightarrow \mY}{{\rm Sym}}\left\{\LP \LCM_\mX\mK, \mY \RP\right\}=0, \quad \forall \mX,\mY\in TM. 
\end{align}
\end{definition} 

\begin{definition} 
A linear self-adjoint mapping $K(\;\cdot\;):TM\rightarrow TM$ is called a Killing mapping if
\begin{align}\label{eq:killing_mapping_equation}
\underset{\mX \leftrightarrow \mY\leftrightarrow \mZ}{{\rm Sym}}\left\{\LP\LCM_\mX K(\mZ),\mY\RP\right\}=0, \quad \forall \mX,\mY,\mZ\in TM,
\end{align} 
where the linear mapping $\LCM_\mX K(\;\cdot\;):TM\rightarrow TM$ is defined as follows
\begin{align}
\LCM_\mX K( \mY)\equiv\LCM_\mX( K( \mY))-K(\LCM_\mX \mY), \quad \forall \mX,\mY\in TM.
\end{align} 
\end{definition} 

One can introduce a Killing tensor as a symmetric form $K(\mX,\mY)=\LP K(\mX),\mY\RP$, which is associated with the conservation law quadratic in momenta. Let $\mathcal K_\alpha$ be a set of $n$ Killing vector fields. Then, one can define the following trivial Killing mapping                  
\begin{align}\label{eq:trivial_tensor}
K(\mX)=\alpha \mX+\sum^n_{\alpha,\beta=1}\gamma^{\alpha\beta}\LP \mX,\mathcal K_\alpha\RP\mathcal K_\beta, \quad \gamma^{\alpha\beta}=\gamma^{\beta\alpha},
\end{align}   
where $\alpha$ and $\gamma^{\alpha\beta}$ is the set of $n(n+1)/2+1$ independent constants in $M$. Note that the trivial Killing mapping does not give new conservation laws. However, one can show the existence of manifolds with nontrivial Killing tensors, which are not associated with the manifold isometries directly.

\section{Generation of a non-trivial Killing tensor}
\label{sec:generation}
Consider a timelike/spacelike {\em foliation} of the manifold $M$ by a smooth family of hypersurfaces $S_\Omega$ parameterized by $\Omega\in\mathbb{R}$ (slices) with the lapse function $\varphi$, and vector field $\xi$ normal to slices ($\LP \xi, \xi \RP\equiv\epsilon=\pm1$). Then, the second fundamental form ${}^\Omega\sigma( \;\cdot\; ,\;\cdot\;):TS\times TS\rightarrow \mathbb R$ and the mean curvature of slices $S_\Omega$ are defined as follows
\begin{align}
    {}^\Omega\sigma(X,Y)\equiv \epsilon\LP \LCM_XY,\xi\RP, \quad
    \forall X,Y\in TS, \qquad
    H\equiv{\rm Tr}({}^\Omega\sigma)/(m-1).
\label{SFF1}
\end{align}

In particular, one can decompose the Killing vector field into the sum $\mK=\mK_\Omega+k_N\xi$, with the normal $k_N\xi$ and tangent $\mK_\Omega\in TS_\Omega$ components. In the general case, the projection $\mathcal K_{\Omega}$ is not a Killing vector in the slices of foliation\cite{Kobialko:2020vqf,Kobialko:2021aqg}. An exception is the case of totally umbilic or totally geodesic slice, where the projection of any Killing field is a conformal or ordinary Killing vector field respectively. Such slices arise if the field generating the foliation is a (conformal) Killing field and/or the spacetime has the structure of a warped/twisted product \cite{Chen}. Therefore, the generation of the Killing vectors in $M$ from Killing vectors $\mathcal{K}_\Omega$ with a nontrivial normal component $k_N$ is possible in the case of the totally geodesic slices only. As we will see further, the case of Killing tensors is more intricate.

Similarly, the Killing mapping can be split to normal and tangent components $K(\; \cdot\;)=K^{(\; \cdot\;)}_\Omega+k^{(\; \cdot\;)}_N\xi$, where $k^{(\; \cdot\;)}_N\xi$ is a normal component and $K^{(\; \cdot\;)}_\Omega\in TS_\Omega$ is a tangent component. In the case of totally geodesic slices, one can lift the Killing tensor from the slice to the whole manifold and obtain a nontrivial normal component $k_N^{(\;\cdot\;)}$. This particular case of totally geodesic slices was considered, for example, in the Ref.\cite{Garfinkle:2010er}. Moreover, if we consider the conformal Killing tensors, a similar technique can be applied in the warped spacetimes \cite{Krtous:2015ona}, where the foliation slices are totally umbilic \cite{Chen}. In this paper, we consider the Killing tensor lift technique for arbitrary slices (not totally geodesic submanifolds). In this case, the second fundamental form ${}^\Omega \sigma$ is not trivial, and Killing equations imply $k^X_N=0$, $K^\xi_{\Omega}=0$. Then, the family of Killing mappings $K_{\Omega}:TS_{\Omega}\rightarrow TS_{\Omega}$ can be lifted from the slices to the Killing mapping $K(\; \cdot\;)=K^{(\; \cdot\;)}_\Omega+k^{(\; \cdot\;)}_N\xi$ in the manifold $M$ with nontrivial normal components, if the following equations hold\cite{Kobialko:2020vqf,Kobialko:2021aqg}
\begin{subequations}
\begin{align}
&k^X_N=0, \quad K^\xi_{\Omega}=0,\quad \xi(k^\xi_N)=0,\quad  X(k^\xi_N)=2 k^\xi_N\LCM_X\ln \varphi-2\LCM_{K^X_{\Omega}}\ln\varphi,\label{eq:kkv1a}\\&\underset{X \leftrightarrow Y}{{\rm Sym}}\{\,\frac{1}{2}
\cdot\LP\LCM_\xi K^X_{\Omega},Y\RP+ \epsilon\cdot {}^\Omega\sigma(X,K^Y_{\Omega})-\epsilon k^\xi_N\cdot{}^\Omega\sigma(X,Y)\}=0.\label{eq:kkv1b}
\end{align}\label{eq:kkv1}
\end{subequations}

Suppose that the manifold $M$ has a collection  $n\leq m-2$ of linearly independent Killing vector fields $\mK_\alpha$ tangent to the slices $S_\Omega$ of the foliation $\mF_\Omega$. Then, such vectors $\mK_\alpha$ are also Killing vectors in the slices $S_\Omega$, and a trivial Killing mapping of the form (\ref{eq:trivial_tensor}) is always defined. Substituting this mapping into equations (\ref{eq:kkv1a}), (\ref{eq:kkv1b}) we obtain
\begin{subequations}
\begin{align}
    &
    X(k^\xi_N)=2 (k^\xi_N-\alpha)\LCM_X\ln \varphi, \quad
    \xi(k^\xi_N)=0, \quad
    X(\alpha)=0, \quad
    X(\gamma^{\alpha\beta})=0,\label{eq:kkv3a}
    \\&
    \label{eq:map_evolution_2}
    2\epsilon(k^\xi_N-\alpha)\cdot{}^\Omega\sigma(X,Y)=\xi(\alpha)\LP X,Y\RP
+\sum^n_{\alpha,\beta=1}\xi(\gamma^{\alpha\beta})\LP X,\mathcal K_\alpha\RP\LP\mathcal K_\beta,Y\RP, 
\end{align}
\label{eq:kkv3b}
\end{subequations}
for any $X,Y\in TS_\Omega$. There is always a trivial solution for these equations
\begin{align}
k^\xi_N=\alpha, \quad \xi(\alpha)=0, \quad \xi(\gamma^{\alpha\beta})=0,
\end{align}
corresponding to the trivial Killing tensor in $M$.  However, in some cases it can also have nontrivial solutions, which corresponds to the nontrivial Killing tensor and new conservation laws. Let us additionally assume that the Gramian matrix $\mG_{\alpha\beta}=\LP\mK_\alpha,\mK_\beta\RP$ is not degenerate ($\mG\equiv \det(\mG_{\alpha\beta}) \neq 0$). Then, we can introduce a basis $\{\mK_\alpha,e_a\}$ in $S_\Omega$ in such a way that $e_a\in\{\mK_\alpha\}^{\perp}$ with $a=1,\ldots,m-n-1$. A non-trivial Killing tensor can be generated using the technique from the following theorem\cite{Kobialko:2021aqg}:

\begin{theorem} 
\label{KBG}
Let the manifold $M$ contains a collection of $n\leq m-2$ Killing vector fields $\mK_\alpha$ with a non-degenerate Gramian $\mG_{\alpha\beta}=\LP\mK_\alpha,\mK_\beta\RP$, tangent to the foliation slices $S_\Omega$ (partially umbilic if $n< m-2$) with the second fundamental form\footnote{The form of the left upper block is a consequence of the tangent Killing vectors, and this does not impose a new condition. The non-diagonal zero elements is a new condition, which is satisfied in many applications. The right lower block is a condition of partially umbilical surfaces, which imposes constraints if ${\rm dim}\{e_{a}\} > 1$. }
\begin{align}\label{eq:second_form_explicit}
{}^\Omega\sigma=\begin{pmatrix}
-\frac{1}{2} \epsilon \cdot \xi\mG_{\alpha\beta} & 0\\
0 & h^\Omega \cdot  \LP e_a,e_b\RP  & 
\end{pmatrix}
\end{align}
Then, there is a nontrivial Killing tensor on manifold $M$, if the following steps can be successfully completed:

{\bf Step one}: Check  compatibility and integrability conditions (\ref{K1}), (\ref{eq:gamma_integrability})
\begin{align}
X(h^{\Omega}\cdot\varphi^3) = 0,
\label{K1}
\end{align}
\begin{align} \label{eq:gamma_integrability}
  X \left(
    \mathcal G^{\alpha\beta}
    -
    \frac{\epsilon}{2 h^{\Omega}} \cdot \xi \mathcal G^{\alpha\beta}
    \right) = 0.
\end{align}

{\bf Step two}: Obtain $\alpha$ from (\ref{K2}) and check the condition (\ref{K2_cond})
\begin{align}
\xi\ln\xi(\alpha)=\xi \ln h^{\Omega}-2\epsilon h^{\Omega},
\label{K2}
\end{align}
\begin{equation}\label{K2_cond}
    X(\alpha)=0.
\end{equation}

{\bf Step three}: Define $\gamma^{\alpha\beta}$ from (\ref{K3n}) using the  conditions $\xi\nu^{\alpha\beta}=0$, $X\gamma^{\alpha\beta}=0$.
\begin{align} \label{K3n}
    \gamma^{\alpha\beta} = 
    \epsilon\frac{\xi(\alpha)}{2 h^{\Omega}} \cdot \mathcal G^{\alpha\beta}
    - \nu^{\alpha\beta},
\end{align}

{\bf Step four}: Using the functions found in the previous steps, construct a Killing map and the corresponding Killing tensor:
\begin{align}
K(\mX)=
\alpha \mX
+ \sum^n_{\alpha,\beta=1}\gamma^{\alpha\beta}\LP \mX,\mathcal K_\alpha\RP\mathcal K_\beta
+ \frac{\xi(\alpha)}{2 h^{\Omega}}\LP\mX,\xi \RP\xi.
\end{align}
\end{theorem}

\section{Connection with photon submanifolds}
\label{sec:photon}

Consider the case of a manifold with two Killing vectors spanning a timelike surface ($\mathcal G<0$). Let us define a Killing vector field $\rho^\alpha$ with index $\alpha=1,2$ numbering Killing vectors of the basis $\{\mathcal{K}_\alpha\}$, which is supposed to have constant components $\rho^\alpha=(\rho,1)$. The quantity $\rho$ can be called the generalized impact parameter (see \cite{Kobialko:2020vqf} for details). However, one can choose arbitrary parametrization of $\rho^\alpha$ up to the norm.

\begin{proposition} 
The fundamental photon surface is a partially umbilic surface with a second fundamental form of the form (\ref{eq:second_form_explicit}), with the following connection between $h^\Omega$ and $\mG_{\alpha\beta}$\cite{Kobialko:2020vqf}
\begin{align}
    \rho^\alpha \mathcal{M}_{\alpha\beta} \rho^\beta = 0,
    \qquad
    \mathcal{M}_{\alpha\beta} \equiv
      \frac{1}{2 h^{\Omega}}\cdot\xi\mG_{\alpha\beta}
    - \mG_{\alpha\beta}
    - \frac{1}{2 h^{\Omega}}\cdot\xi\ln \mathcal G\cdot\mG_{\alpha\beta}.
\label{FPS3}
\end{align}
\end{proposition}

If the surface under consideration is totally umbilic $\mathcal{M}_{\alpha\beta}=0$, it is obviously a fundamental photon surfaces for any $\rho$. Since totally umbilic surfaces usually exist in spherically symmetric solutions (both static and non-static) or non-rotating solutions with NUT-charge \cite{Galtsov:2019bty}, and they have been considered in detail in a number of works \cite{Claudel:2000yi,Koga:2020akc}, we will focus on the case $\mathcal{M}_{\alpha\beta}\neq0$. 

Consider the foliation generating a nontrivial Killing tensor in accordance with Theorem \ref{KBG}, and ask the question  whether its slice $S_\Omega$ is a fundamental photon surface. First of all, we need to solve the quadratic equation (\ref{FPS3}) for $\rho$ and check the condition $\rho^\alpha \mathcal{G}_{\alpha\beta} \rho^\beta\geq0$. It has nontrivial solution if the eigenvalues of the matrix $\mathcal{M}_{\alpha\beta}$ have different signs, that is $\mathcal M\equiv\det(\mathcal M_{\alpha\beta})<0$. Then the solution for $\rho$ reads as
\begin{align}\label{PR1a}
    \rho = \frac{-\mathcal M_{12}\pm \sqrt{-\mathcal M}}{\mathcal M_{11}}.
\end{align}
Condition $\rho^\alpha \mathcal{G}_{\alpha\beta} \rho^\beta\geq0$ is satisfied if the following inequality holds
\begin{align}\label{PR1b}
\pm 2(\mathcal G_{12}\mathcal M_{11}-\mathcal G_{11}\mathcal M_{12})\sqrt{-\mathcal M}-2\mathcal G_{11} \cdot \mathcal M+\mathcal M_{11} \cdot\mathcal G \cdot {\rm Tr}(\mathcal M)\geq0,
\end{align}
where ${\rm Tr}(\mathcal M)=\mathcal M_{\alpha\beta} \mathcal G^{\alpha\beta}=-2-(2h^\Omega)^{-1}\cdot\xi\ln\mG$. Equation (\ref{PR1b}) defines the so-called photon region \cite{Grenzebach:2014fha,Grenzebach:2015oea}, which arises as a flow of fundamental photon surfaces \cite{Kobialko:2020vqf}. However, it has not been proven that the expression (\ref{PR1a}) for $\rho$ is constant in every slice. But the integrability condition (\ref{eq:gamma_integrability}) guarantees that it is true in fact \cite{Kobialko:2021aqg}. Therefore, we have the following theorem.

\begin{theorem} 
Let $S_\Omega$ be a non totally umbilic foliation slice with compact spatial section satisfying all conditions of the theorem \ref{KBG} for $\text{dim} \{\mathcal{K}_\alpha\}=2$. Then maximal subdomain $U_{PS} \subseteq S_\Omega$ such that the inequality (\ref{PR1b}) holds for all $p\in U_{PS}$ is a fundamental photon surface\footnote{In the case of not compact spatial section, the slice is not a fundamental photon surface by definition \cite{Kobialko:2020vqf}. However, the theorem can be generalized for such not compact surfaces.}.
\end{theorem}
 
In particular, the region $U_{PR}\subseteq M$, such that the inequality (\ref{PR1b}) holds for any point $p\in U_{PR}$, is a photon region. This theorem generalizes the connection between the existence of Killing tensors of this type and photon surfaces or spherical null geodesics, which was noted in Refs. \cite{Koga:2020akc,Pappas:2018opz,Glampedakis:2018blj}. Unfortunately, in the opposite direction, the theorem is not fair, since the existence of fundamental photon surfaces does not guarantee the existence of the Killing tensor. As a counterexample, one can suggest Zipoy-Voorhees metric \cite{Kodama:2003ch} where the fundamental photon surfaces exists \cite{Galtsov:2019fzq} but there is no nontrivial Killing tensor \cite{Lukes-Gerakopoulos:2012qpc}. Nevertheless, the existence of fundamental photon surfaces can serve as a sign that the Killing tensor can be presented in the corresponding metric, and it is advisable to check the conditions of consistency and integrability.

\section{Axially symmetric spacetimes} \label{sec:axially} 
 
Consider a Lorentzian manifold $M$ with the metric tensor
\begin{align}
ds^2=-f (dt-\omega d\phi)^2+\lambda dr^2+ \beta d\theta^2 +\gamma d\phi^2. 
\end{align}
where all metric components depend on $r$ and $\theta$ only and the foliation with timelike slices $r=\Omega$. Generally, this metric possesses two Killing vectors $\mathcal{K}_1= \partial_t$, $\mathcal{K}_2 = \partial_\varphi$. One can find that the second fundamental form of these slices has the form (\ref{eq:second_form_explicit}) and other quantities are
\begin{align}
    \xi = \lambda^{-1/2}\partial_r, \quad
    h^{\Omega} = -\frac{1}{2}\lambda^{-1/2}\cdot\partial_r \ln \beta, \quad
    \varphi=\lambda^{1/2},\quad
    \mathcal G^{\alpha\beta} = \frac{1}{\gamma}\begin{pmatrix}
        \omega^2- \gamma f^{-1}  & \omega  \\
        \omega  & 1 \\
    \end{pmatrix}.
\end{align}
In this case, the number of Killing vector is one less than the slices dimension, so the boundary $n \geq m - 2$ saturates and the partially umbilic condition just imposes a relation on $h^\Omega$. The compatibility and integrability conditions (\ref{K1}), (\ref{eq:gamma_integrability}) take the form
\begin{align}
    \partial_\theta(\lambda \cdot\partial_r\ln\beta)=0,\qquad
    \partial_\theta \left(
        \mathcal G^{\alpha\beta}
        +
        \frac{1}{\partial_r \ln \beta} \partial_r \mathcal G^{\alpha\beta}
    \right) = 0.
\end{align}
The Eq. (\ref{K2}) can be solved as follows
\begin{align}
    \alpha=A_\theta \cdot\beta+B_\theta,
\end{align}
where the arbitrary functions $A_\theta$, $B_\theta$ depend on $\theta$ only, obeying the condition $\partial_\theta \alpha = 0$. As the result, we have one more necessary condition for the case in this section: the function $\beta$ must be of the form 
\begin{equation}\label{eq:beta_condition}
    \beta(r,\theta) = \beta_1(\theta) \beta_2(r) + \beta_3(\theta),
\end{equation}
where $\beta_{1,2,3}$ are some functions of the corresponding variables. In particular, the normal component is $k^\xi_N=B_\theta$. Next, we can define the matrix $\gamma$:
\begin{align}
    \gamma^{\alpha\beta} = 
    - \beta A_\theta \cdot \mathcal G^{\alpha\beta}
    - \nu^{\alpha\beta}.
\end{align}
The integrability condition guarantees that $\gamma^{\alpha\beta}$ always satisfies the equations (\ref{eq:kkv3b}) for some $\nu^{\alpha\beta}$ depending only on $\theta$. On the other hand, we have to find a $\nu^{\alpha \beta}$ that makes the equation $\partial_r \gamma^{\alpha\beta}= 0$ true. Therefore, we can omit the $\theta$-dependent part in $\gamma^{\alpha\beta}$ to some constant matrix instead of looking for $\nu^{\alpha\beta}$. Combining everything together, we get the final Killing tensor in the holonomic basis 
\begin{align}
    K^{\mu\nu} =
    \alpha g^{\mu\nu}
    + \sum_{\alpha,\beta=t,\phi}\gamma^{\alpha\beta} \mathcal{K}^\mu_\alpha \mathcal{K}^\nu_\beta
    - \beta A_\theta \lambda^{-1} \delta_r^\mu \delta_r^\nu.
\end{align}

The compatibility and integrability conditions, as well as the condition on the function $\beta$, are invariant under the multiplicative transformations of the form
\begin{align}
    \lambda\rightarrow\lambda'=u(r)\lambda, \quad \beta\rightarrow\beta'=v(\theta)\beta.
\end{align}
If $\beta$ possesses the aforementioned form (\ref{eq:beta_condition}), one can simplify the integrability condition by the substitution $\mathcal{G}^{\alpha\beta} = \tilde{\mathcal{G}}^{\alpha\beta} \cdot \beta_1 / \beta$. Then, the integrability condition is $\partial_\theta\partial_r \tilde{\mathcal{G}}^{\alpha\beta} = 0$, which is solved by $\tilde{\mathcal{G}}^{\alpha\beta} = \tilde{\mathcal{G}}_r^{\alpha\beta}(r) + \tilde{\mathcal{G}}_\theta^{\alpha\beta}(\theta)$. This generalizes the result of Ref. \cite{Johannsen:2013szh}, where the similar condition was obtained from the separability of the Hamilton-Jacobi equation. In our case, we have also included the $\beta_1(\theta)$ term. Furthermore, the compatibility condition and the function form (\ref{eq:beta_condition}) leads to the form of $\lambda=\lambda_r(r) \beta/\beta_1$, where $\lambda_r$ is an arbitrary function of $r$.

\subsection{Kerr metric}

As a simple illustration, consider   Kerr solution in the Boyer-Lindquist coordinates:
\begin{equation}\label{Kerr}
    ds^2 =
    - \frac{\Delta - a^2\sin^2\theta}{\Sigma}(dt - \omega d\phi)^2
    +\Sigma \left(
          \frac{dr^2}{\Delta}
        + d\theta^2
        + \frac{\Delta \sin^2\theta}{\Delta - a^2\sin^2\theta} d\phi^2
    \right),
\end{equation}\
\begin{subequations}
\begin{equation}
    \Sigma = r^2 + a^2 \cos^2\theta,\qquad
    \omega=\frac{-2Mar\sin^2\theta}{\Delta-a^2\sin^2\theta},
\end{equation}
\begin{equation}
    \Delta = r(r-2M) + a^2.
\end{equation}
\end{subequations}
In the Kerr metric, $\beta=r^2+a^2\cos^2\theta$, $\lambda=\beta/\Delta$,   satisfy the compatibility condition. One can explicitly verify that $\mathcal{G}^{\alpha\beta}$ satisfies the integrability equation. In this case $\alpha=r^2$, $A_\theta=1$ and $k_N^\xi=B_\theta=-a^2\cos^2\theta$ (here we have fixed the multiplicative integration constant, which appears due to the linearity of Killing equations). The part of $\gamma^{\alpha\beta}$ independent on $\theta$ reads
\begin{align}
\gamma^{\alpha\beta}=\Delta^{-1}\left(\begin{matrix}
 (a^2+r^2)^2 &   a(a^2+r^2) \\
a(a^2+r^2) &  a^2 \\
\end{matrix}\right).
\end{align}
Finally, we get $\alpha$ and $\gamma^{\alpha\beta}$, which correspond to the well-known nontrivial Killing tensor for Kerr solution
\begin{align}
    &
    K^{\mu\nu} =
    r^2 g^{\mu\nu}
    +\Delta^{-1} S^\mu S^\nu
    - \Delta \delta_r^\mu \delta_r^\nu,
    \\\nonumber &
    S^\mu = s \delta^\mu_t + a \delta^\mu_\varphi,\qquad
    s = r^2+ a^2.
\end{align}

\subsection{  Plebanski-Demianski solution}
This is a less trivial example of the type D solution of Einstein-Maxwell equations with the cosmological constant, which contains also an acceleration parameter.   The metric line element $ds^2$  is more conveniently presented in the conformally related frame using the Boyer-Lindquist coordinates: 
\begin{align}
\Omega^2ds^2&=\Sigma\left(\frac{dr^2}{\Delta_r}+\frac{d\theta^2}{\Delta_\theta}\right)+\frac{1}{\Sigma}
\left((\Sigma+a\chi)^2\Delta_\theta\sin^2\theta-\Delta_r\chi^2\right)d\phi^2 \\
\label{Sol222}
&+\frac{2}{\Sigma}\left(\Delta_r\chi-a(\Sigma+a\chi)\Delta_\theta\sin^2\theta\right)dt d\phi-\frac{1}{\Sigma}
\left(\Delta_r-a^2\Delta_\theta\sin^2\theta \right)dt^2,
\end{align}
where we have defined the functions
\begin{align}
\Delta_\theta &=1-a_1\cos\theta-a_2\cos^2\theta, \qquad \Delta_r=b_0+b_1r+b_2r^2+b_3r^3+b_4r^4\,,\\
\Omega &=1-\lambda(N+a \cos\theta)r, \quad \Sigma=r^2+(N+a\cos\theta)^2\,,\quad
\chi =a \sin^2\theta-2N(\cos \theta+C)\,,
\end{align}
with the following constant coefficients in $\Delta_\theta$ and $\Delta_r$:
\begin{align}
a_1 &=2aM\lambda-4aN\left(\lambda^2(k+\beta)+\frac{\Lambda}{3}\right), \quad a_2=-a^2\left(\lambda^2(k+\beta)+\frac{\Lambda}{3}\right),\quad
b_0=k+\beta, \\  
b_1 &=-2M,\quad
b_2=\frac{k}{a^2-N^2}+4MN\lambda-(a^2+3N^2)\left(\lambda^2(k+\beta)+\frac{\Lambda}{3}\right),\\
b_3 &=-2\lambda\left(\frac{kN}{a^2-N^2}-(a^2-N^2)\left(M\lambda-N\left(\lambda^2(k+\beta)+\frac{\Lambda}{3}\right)\right)\right),\\
b_4 &=-\left(\lambda^2k+\frac{\Lambda}{3}\right),\\ 
k &=\frac{1+2MN  \lambda-3N^2\left(\lambda^2\beta+\frac{\Lambda}{3}\right)}{1+3\lambda^2N^2(a^2-N^2)}(a^2-N^2), \quad
\lambda=\frac{\alpha}{\omega}, \quad \omega=\sqrt{a^2+N^2}\,.
\end{align}
Generally, the coordinates $t$ and $r$ range over the whole $\mathbb R$, while $\theta$ and $\phi$ are the
standard coordinates on the unit two-sphere. Seven independent parameters $M,N,a,\alpha,\beta,\Lambda,C$ can be physically interpreted as  mass,  NUT parameter (magnetic mass),  rotation parameter,  acceleration parameter, $\beta=e^2+g^2$ comprises the electric $e$ and magnetic $g$ charges,  $\Lambda$ is the cosmological constant, and the constant $C$ defines the location of the Misner string. 
 
{\bf The first step} is to check the compatibility and integrability conditions (\ref{K1}), (\ref{eq:gamma_integrability}). The first one holds if $\alpha \cdot a=0$, i.e. either the acceleration $\alpha$ or the rotation $a$ are zero. Indeed, as  shown in Ref. \cite{Kubiznak:2007kh}, the general PD solution with acceleration possesses a conformal Killing tensor, but not the usual one. In the case $a=0$, the second condition does not hold. So, further we will consider the solution with zero acceleration $\alpha=0$, which corresponds to the dyonic Kerr-Newman-NUT-AdS solution.

In the {\bf second step} we pick up the $r$-dependent part from $\beta = \Sigma / \Delta_\theta$ for $\alpha$
\begin{align}
\beta=\frac{\Sigma}{\Delta_\theta} \quad \Rightarrow \quad \alpha=r^2, \quad A_\theta=\Delta_\theta.
\end{align}
Similarly, as the {\bf third step}, the $r$-dependent part for $\gamma^{\alpha\beta}$ is defined as
\begin{align}
\gamma^{\alpha\beta} = 
\Delta^{-1}_r \begin{pmatrix}
     s^2  & as  \\
      as  & a^2 \\
\end{pmatrix}, \quad s = \Sigma + a\chi = r^2 + a^2 - 2 a C N + N^2.
\end{align}
In the last {\bf fourth step}, we obtain the nontrivial Killing tensor for the Kerr-Newman-NUT-AdS metric:
\begin{align}
    &
    K^{\mu\nu} =
    r^2 g^{\mu\nu}
    +\Delta^{-1}_r S^\mu S^\nu
    - \Delta_r \delta_r^\mu \delta_r^\nu, \quad
    S^\mu = s \delta^\mu_t + a \delta^\mu_\varphi.
\end{align}
\section{Gal'tsov-Kechkin (GK) solution}
In 1994 one of the present authors in collaboration with O. Kechkin derived the general stationary charged black hole solution within the Einstein-Maxwell- dilaton-axion (EMDA) gravity, which is the ${\cal N} =4, D=4$ supergravity consistently truncated to the theory with one vector field \cite{Galtsov:1994pd}. This solution was seven-parametric, containing   mass $M$,  electric and magnetic charges $Q,P$,   rotation parameter $a$,   NUT $N$ and   asymptotic values of the dilaton and axion (irrelevant for the metric) as independent parameters. Less general (without NUT) solution  was derived by A. Sen \cite{Sen:1992ua} in the context of the dimensionally reduced effective action of the heterotic string,  and now it is commonly referred as Kerr-Sen metric. Non-rotating solutions with NUT were independently obtained by Kallosh et al.\cite{Kallosh:1994ba}  and Johnson and Myers\cite{Johnson:1994nj}. 
Now the Kerr-Sen metric is often considered as a deformed Kerr solution in modeling deviations from the standard picture of  black holes. 

The GK solution \cite{Galtsov:1994pd}  was shown \cite{Garcia:1995qz} to belong to type Petrov I, contrary to Kerr-Newman-NUT solution in the Einstein-Maxwell gravity. The same was shown for Kerr-Sen solution without NUT \cite{Burinskii:1995hk}. Though the Kerr-Sen metric is not type D, the Hamilton-Jacoby equation was shown to be separable for it  \cite{Konoplya:2018arm,Lan:2018lyj}, though the Killing tensor was not found explicitly (for further discussion see \cite{Papadopoulos:2020kxu}). But for the solution with NUT,   it was claimed that no second rank Killing tensor exists \cite{Siahaan:2019kbw}. So here we use  the new technique to resolve this controversy. 
 
The line element of the GK solution can be written  in the form (\ref{Kerr})
where the functions $\Delta$, $\omega$, $\Sigma$ are redefined as follows
\begin{align}
   & \Delta = (r - r_{-}) (r - 2M) + a^2 - (N-N_{-})^2, \n\\
   & \Sigma = r(r-r_{-}) + (a\cos\theta + N)^2 - N_{-}^2, \n\\
    &\omega = \frac{-2w}{\Delta - a^2 \sin^2\theta},\quad
     w =  N \Delta \cos\theta
        + a \sin^2\theta \left( M(r-r_{-}) + N(N - N_{-}) \right)\n .
\end{align}
The full solution also contains Maxwell, dilaton $\phi$ and axion $\kappa$ fields (whose form can be found in \cite{Galtsov:1994pd}), and represents a family with seven parameters: mass $M$, NUT charge $N$, rotation parameter $a$, electric and magnetic charges $Q$ and $P$, and asymptotic complex axidilaton charge $z_\infty = \kappa_\infty + i \e^{-2\phi_\infty}$, irrelevant for the metric.
The following abbreviations are used:
\begin{equation}
    r_{-} = \frac{M |\mathcal{Q}|^2 }{|\mathcal{M}|^2},\quad
    N_{-} = \frac{N |\mathcal{Q}|^2 }{2|\mathcal{M}|^2},\quad
    \mathcal{M} = M + i N,\quad
    \mathcal{Q} = Q - i P,\quad
    \mathcal{D} = - \frac{{\mathcal{Q}^*}^2}{2\mathcal{M}}.
\end{equation}
The metric is presented in the Kerr-like form, but the metric functions are essentially different. The solution  reduces to Kerr-NUT for $Q=P=0$.  

Now we apply our procedure to construct the Killing tensor.
It can be easily verified that the consistency (\ref{K1}) and the integrability (\ref{eq:gamma_integrability}) conditions are satisfied. The uplift turns out to be as simple as in the vacuum Kerr example, leading to the result   
\begin{align}
    &
    K^{\mu\nu} =
    r(r-r_{-}) g^{\mu\nu}
    +\Delta^{-1} S^\mu S^\nu
    - \Delta \delta_r^\mu \delta_r^\nu,
    \\\nonumber &
    S^\mu = s \delta^\mu_t + a \delta^\mu_\varphi,\qquad
    s = r(r-r_{-}) + a^2 + N^2 - N_{-}^2.
\end{align}
This expression is new and is applicable both to the Kerr-Sen solution $N=N_-=0$, and for the full GK solution.

\section{Conclusions}
In this article, we reviewed new geometric method of generating  Killing tensor in spacetimes with codimension one  foliation \cite{Kobialko:2021aqg}. Using our general lift equations one can try to rise a trivial Killing tensor defined in the slices into a nontrivial Killing tensor in the bulk. For this, the foliation must satisfy some consistency and integrability  conditions, which we have presented explicitly. Furthermore, we have completely solved the lifting equations in terms of the functions $\alpha$, $\gamma^{\alpha\beta}$ and formulated the theorem (\ref{KBG}) for generation of the non-trivial Killing tensor.

Finding a foliation compatible with integrability and consistency conditions can be a challenging task. The existence of such a foliation means that the slices represent fundamental photon surfaces provided the corresponding inequalities hold. This generalizes the result of Ref. \cite{Koga:2020akc} to the case of general stationary spaces. Conversely, the existence of fundamental photon surfaces, though does not guarantee the existence of the Killing tensor, but may serve as an indication that such tensor may exist. Therefore, it is recommended to check the consistency and integrability conditions for fundamental photon surfaces, if these are known. This property makes the search for fundamental photon surfaces important also for studying the integrability of geodesic motion. It is tempting to conjecture that the existence of fundamental photon surfaces implies the existence of Killing tensor if the slice is equipotential or spherical \cite{Cederbaum:2019rbv}.

Using this technique, we were able to derive Killing tensor for EMDA dyon with NUT (GK solution \cite{Galtsov:1994pd}), which is also valid for Kerr-Sen solution as a particular case.
\section*{Acknowledgements} 
The work was supported by the Russian Foundation for Basic Research on the project 20-52-18012Bulg-a, and the Scientific and Educational School of Moscow State University “Fundamental and Applied Space Research”. I.B. is also grateful to the Foundation for the Advancement of Theoretical Physics and Mathematics ``BASIS'' for support.

\end{document}